# ONSET OF REPTATIONS AND CRITICAL HYSTERETIC BEHAVIOR IN DISORDERED SYSTEMS

O. Hovorka, G. Friedman[1]
Electrical and Computer Engineering
Magnetic Microsystems Laboratory
Drexel University

**Abstract.**
**Zero-temperature random coercivity Ising model with antiferromagnetic-like interactions is used to study closure of minor hysteresis loops and wiping-out property in hysteretic behavior. Numerical simulations in two dimensions as well as mean-field modeling show a critical phenomenon in the hysteretic behavior associated the loss of minor loop closure and the onset of reptations. Power law scaling of the extent of minor loop reptations is observed.**

Disordered systems having many metastable states can evolve irreversibly even when the external driving force changes slowly. This irreversible process is called hysteresis and is observed in a variety of systems of different physical origins (magnets, superconductors, ferroelectrics, shape memory alloys, porous media, biological and social systems and many others). Cyclic variations of the external driving force result in formation of closed hysteresis loops in most cases. Often these hysteresis loops are observed to close immediately at the end of the very first cycle. Such behavior has been often assumed to occur in hysteretic systems and various terms, such as return point memory (RPM) [1] and wiping-out [2], have been associated with the immediate closure of minor hysteresis loops. On the other hand, absence of RPM has also been observed [3-6], particularly at the level of the microscopic system state. Often gradual rather than immediate stabilization of minor hysteresis loops is noted. Various terms have been used to refer to this gradual minor loop closure. The term *reptation* (Neél [7, 8]) will be used here.

Analysis carried out in this paper demonstrates that an abrupt transition (similar to equilibrium phase transitions) from the immediate closure of minor hysteresis loops to reptations can be induced in disordered systems by tuning the disorder or interactions within the system. The system analyzed in this paper is similar to the Random Field Ising Model (RFIM). Previous work based on zero-temperature RFIM with cooperative (ferromagnetic-like) interactions carried out by Jim Sethna, Karin Dahmen and their co-workers [9] had revealed the possibility of critical phenomena in the hysteresis process associated with a sudden clustering of Barkhausen avalanches induced by tuning the disorder. They also found an elegant proof that absence of competitive (antiferromagnetic-like) interactions results in return point memory (RPM) not only for average quantities, but also for the microscopic state (spin state) of the system.

In this paper we also model hysteretic processes using a collection of interacting spins, except that each spin displays elementary hysteretic behavior of a bi-stable switch with symmetrical "up" and "down" switching thresholds. These bi-stable spins are assumed to compete with each other through antiferromagnetic-like interactions. The magnitude of the switching thresholds for each bi-stable spin is chosen randomly and, for this reason, the model can be called Random Coercivity Ising Model (RCIM). In this work mean-field analysis and numerical simulations of RCIM are employed to show that a critical phenomenon in the hysteresis process is induced when competitive interactions between the spins become sufficiently strong. It will also be shown that this critical phenomenon is directly associated with loss of the return point memory (RPM) and onset of reptations.

---
[1] Send comments to gary@ece.drexel.edu

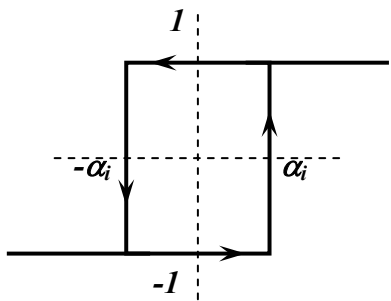

Figure 1 – hysteresis loop characterizing elementary switch $\hat{\gamma}_{\alpha_i}$ with symmetrical thresholds $-\alpha_i$ and $\alpha_i$

The state of the RCIM is the collective state of bi-stable switches $s_i$ each characterized by a rectangular hysteresis loop $\hat{\gamma}_{\alpha_i}$ with symmetrical switching thresholds $\alpha_i$ and $-\alpha_i$. Behavior of each switch is illustrated in Figure 1. The switching of these bi-stable devices at time $n$ is described by the formula:

$$s_i(F(n)) = \hat{\gamma}_{\alpha_i}(F(n)) = \begin{cases} 1, & \text{if } F(n) > \alpha_i \\ \hat{\gamma}_{\alpha_i}(F(n-1)), & \text{otherwise} \\ -1, & \text{if } F(n) < -\alpha_i \end{cases} \quad (1a)$$

where $F$ is the effective force acting on a given switch. Different switches $i$ and $j$ compete with each other and $J_{ij}$ characterizes the strength of their interactions. As the external force $h$ evolves in time, the state of the RCIM evolves by Glauber dynamics (spins are flipped one at a time) according to:

$$s_i(n+1) = \hat{\gamma}_{\alpha_i}(F(n)) = \hat{\gamma}_{\alpha_i}(h(n) - \sum_{j \neq i} J_{ij} s_j(n)) \quad (1b)$$

In general the state of the RCIM at any time depends on the rate of variation of the external force $h$. However, since we are interested in a rate-independent hysteresis, the external force in our analysis is assumed to change adiabatically in small steps, while the state is allowed to equilibrate after each step (spins align along their local fields) and before the external force changes again. The size of the external force step is chosen each time to be no larger than that required to begin switching one spin. Such evolution regime has been called adiabatic.

The model described above can be viewed as a prototype for some real physical systems. One example is a system consisting of nano-magnets. Such systems have been experimentally constructed using lithography and self-assembled templates [10,11] and have been investigated for applications in magnetic recording. The individual nano-magnets, when isolated from each other, behave as bi-stable switches due to inherent anisotropy and strong ferromagnetic exchange interactions spins within each magnet. On the other hand, different nano-magnets interact magnetostatically tending to oppose each other's magnetization and promoting competition. Another example is the so-called AFC (antiferromagnetically coupled) recording media frequently employed today in magnetic storage applications.

Numerical simulations on systems consisting of up to $10^5$ different bi-stable switches positioned on a square 2D lattice have been carried out. The switching thresholds for the individual switches are randomly generated (using Gaussian random generator) at the beginning of each simulation mimicking the effects of quenched-in disorder. In these initial simulations only the nearest neighbor interactions are assumed to exist in order to speed up the simulations. The strength $J_{ij} = J$ of these interactions is the same throughout the system. Figure 2 exemplifies the results showing that return point memory holds as long as the strength of the interactions is below critical. Above the critical interaction strength return point memory gradually disappears and minor loops do not close at their reversal point at the end of the first cycle.

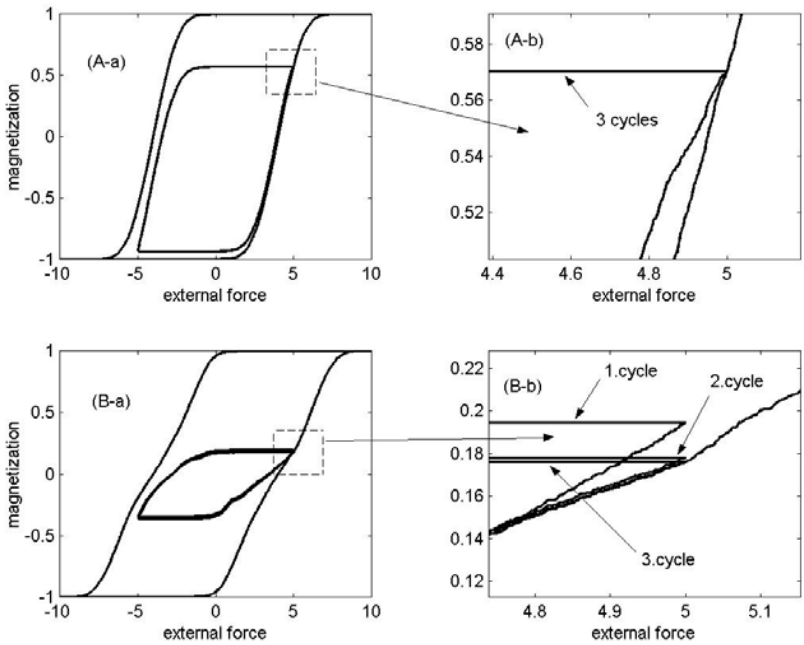

*Figure 2 – Illustration of minor loop having the same reversal values repeated three times for different strength of interaction. (A-a) Minor loop closes at the end of the first cycle showing RPM. Magnetization state follows the same path for each cycle. In this case strength of interaction is below critical. (A-b) Magnification of the minor loop near the reversal point. (B-a) Non-closure and reptation of the minor loops with the same reversal values occurs for the interaction above critical. (B-b) Magnification of the minor loop near the reversal point shows non-closure and reptation.*

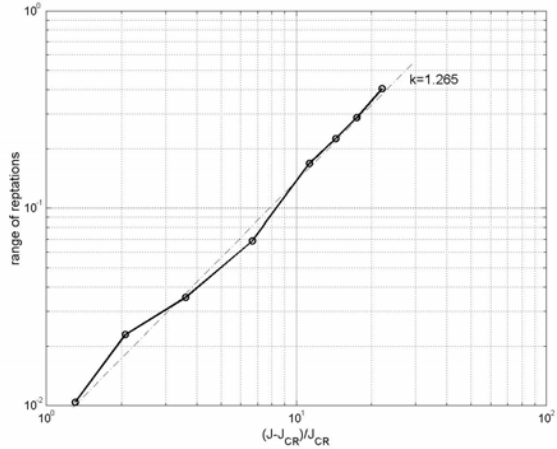

*Figure 3. Logarithmic plot of the dependence of range of reptation on tuning parameter $(J - J_C)/J_C$. Dotted line is a power law fitting indicating the exponent of approximately 1.265.*

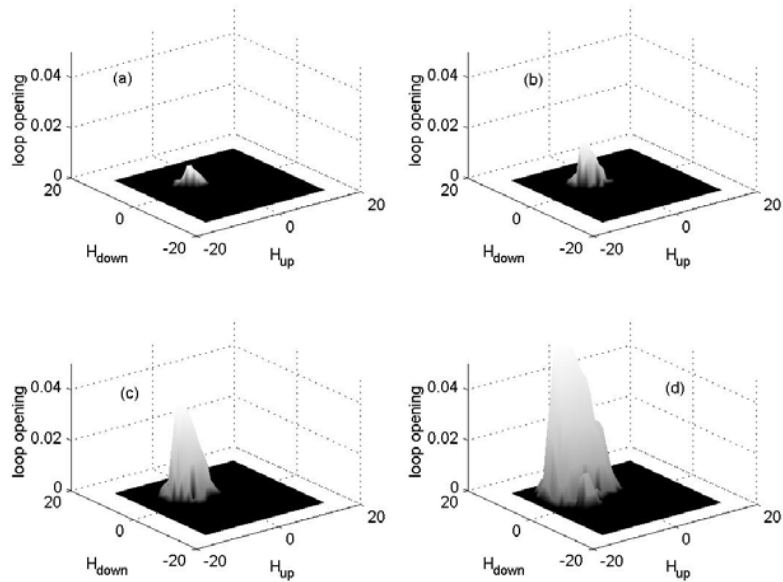

*Figure 4. A sequence of loop opening functions (loop opening as function of the two reversal values) numerically calculated for different interaction strengths. a) Interaction slightly exceeding critical results in very few loops that have a small opening at the end of the first cycle; b) The opening of the few open loops increases quickly as the interaction strength increases; c) and d) Both, the loop opening and the range of reversal values over which open loops are found increase as the interaction strength increases further.*

Several criteria can be chosen to characterize the extent of reptations in the system. One of these criteria is the range of the minor loop reversal values for which reptations are observed. We will call this the *range of reptations*. To define the range of reptations, consider minor loops that are attached to the increasing branch of the major hysteresis loop. In this case, a pair of external force reversal values, a lower and an upper one, completely identifies a minor loop. This minor loop corresponds, therefore, to a point on a plane spanned by the two loop reversal variables. A range of reversal values for which the minor loops do not close immediately corresponds to some area on this plane. This area is what we call the *range of reptations*. The range of reptations can be calculated by meshing the plane of the reversal variables and simulating the minor loops for each discrete point on this mesh. Sufficiently fine mesh size was chosen in simulations by making sure that the range of reptations calculated for each RCIM realization did not change with further mesh refinement. Figure 3 demonstrates how the range of reptation varies as the strength of the antiferromagnetic nearest neighbor interaction $J$ is varied. The range of reptation below the critical interaction strength $J_c$ is not shown in this figure because it is zero (all minor loops close and RPM holds). It is worth noting that above the critical interaction strength $J_c$ the scaling of the range of reptation with the tuning parameter $r = (J - J_C)/J_C$ can be well fit with a power law.

Another important characteristic of the extent to which RPM is violated is the magnitude of the minor loop openings at the end of the first cycle. Obviously the magnitude of the loop opening depends on the choice of the two reversal values of the loop. The loop opening as a function of the two reversal values and the variation of the loop opening function with the interaction strength is illustrated in Figure 4. Integral of the loop opening function over the entire *range of reptations* will be called the *extent of loop opening*. Figure 5 demonstrates that scaling of the *extent of loop opening* with the tuning parameter $r = \dfrac{(J - J_c)}{J_c}$ can also be well fit with a power law.

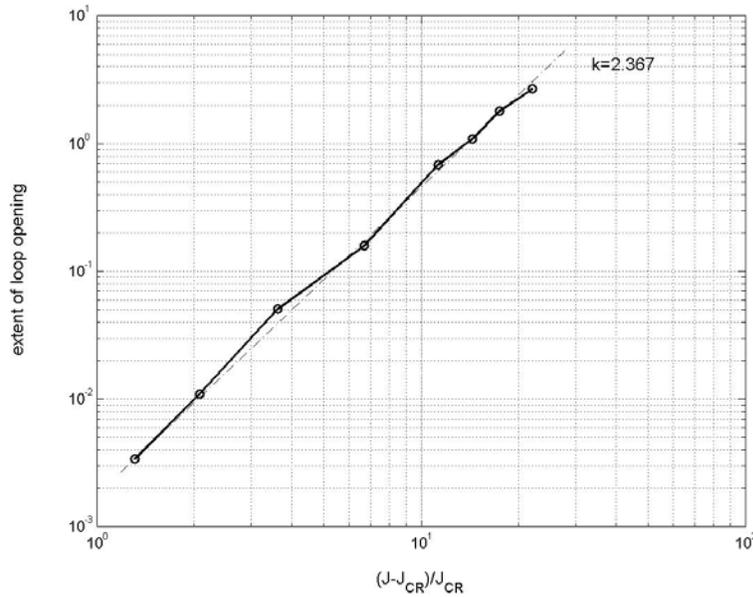

*Figure 5. Logarithmic plot of the dependence of the extent of loop opening on tuning parameter $(J - J_C)/J_C$. Dotted line is a power law fitting indicating the exponent of approximately 2.367.*

Abrupt transition from the hysteretic behavior described by RPM to one associated with reptations as well as power law scaling of reptation characteristics with the tuning parameter suggests the existence of a critical phenomenon that is similar to an equilibrium phase transition. Given the fact that the RCIM considered here is based on antiferromagnetic interactions, it is reasonable to investigate the behavior of the system property that is analogous to the order parameter of the classical antiferromagnetic Ising model. In the classical antiferromagnetic Ising model [12], one considers a lattice of spins as consisting of two interpenetrating sub-lattices. Within each sub-lattice the spins do not interact with each other and antiferromagnetic interactions occur only between neighboring sites on the two sub-lattices. Average spin state (magnetization) can be calculated separately for each of the two sub-lattices and the difference between these average spin states is the order parameter of this Ising model. Below some critical temperature (frequently called the Neél temperature), this order parameter is non-zero and has inversion symmetry. Above this critical temperature the order parameter becomes zero.

By analogy we also split the entire lattice of the bi-stable switches into two interpenetrating sub-lattices for our 2D RCIM system. We computed average spin state for each sub-lattice and looked at the difference between them as the system was cycled around the major hysteresis loop for different strengths of antiferromagnetic interactions. Figure 6 illustrates how the difference of the average spin state between the two sub-lattices behaves when the RCIM (fixed realization of disorder) evolves along the increasing branch of the major hysteresis loop. For low interaction strength the difference of average spin states on the two sub-lattices oscillates in a random fashion as the external force increases. Above some interaction strength, a clear difference in the average spin states emerges near the coercive field of the RCIM. Thus, it appears that the 2D RCIM model does display a critical phenomenon similar to the phase transition of the classical antiferromagnetic Ising model. Moreover, when we compared the interaction strength at which a clear difference between the average spin states emerged with the critical interaction strength at which the onset of reptations occurred for the RCIM with a given disorder realization, we found that these were nearly the same. However, this comparison could not be made more rigorously in our current set up because deciding at which interaction strength the difference in the spin average stopped oscillating required a judgment call.

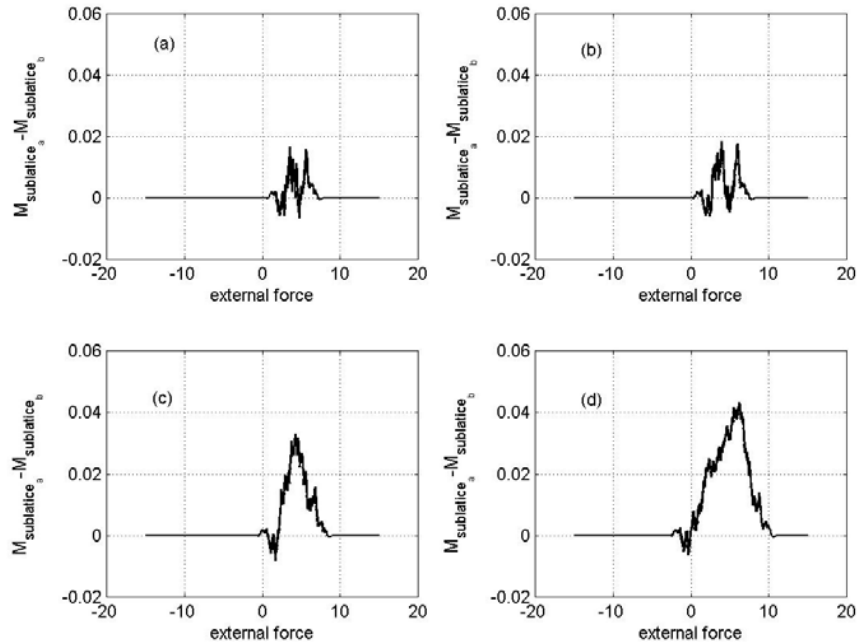

*Figure 6 – Behavior of the difference of average spin state between the two sublattices at different tuning parameters; a) r ~ - 0.94, b) r ~ - 0.4, c) r ~ 1 and d) r ~ 4.3. Figures were plotted at the same scale for better illustration.*

Numerical simulations described above suggest, therefore, that a critical phenomenon in the hysteretic behavior of the RCIM occurs when one tunes the strength of the antiferromagnetic interactions relative to the magnitude of the disorder. It appears that a non-zero difference between average spins of the two interpenetrating lattices is induced as a result of this critical phenomenon and that this coincides with the loss of RPM and the onset of reptations. However, a relatively small size of the simulated systems (up to $10^5$) does not allow us to be absolutely certain of these conclusions. While we did notice that the critical interaction strength increases with the increasing system size, it remains to be confirmed that this increase is bounded. We hope to be able to report more about such simulations in the near future.

To demonstrate that critical transition from hysteresis with RPM to hysteresis with reptations can be real and is not just an artifact in our simulations we employ mean-field modelling. Our mean-field model mimics Neel's mean-field model for antiferromagnets [12]. The entire collection of the bi-stable switches is split into two sub-groups. Within each sub-group the bi-stable switches are non-interacting. However, the effective force responsible for switching within one sub-group includes a contribution proportional to the average state of the other sub-group.

$$\mu_a(n) = \frac{1}{N}\sum_{i \in a} \hat{\gamma}_{\alpha_i}\big(h(n-1) - J\mu_b(n-1)\big) = \hat{P}\big[h(n-1) - J\mu_b(n-1)\big] \quad (2a)$$

$$\mu_b(n) = \frac{1}{N}\sum_{i \in b} \hat{\gamma}_{\alpha_i}\big(h(n-1) - J\mu_a(n-1)\big) = \hat{P}\big[h(n-1) - J\mu_a(n-1)\big] \quad (2b)$$

Operator $\hat{P}$ above is a hysteresis operator whose output can be viewed as the average of the outputs of independent bi-stable switches. This operator belongs to a class of hysteresis operators known as Preisach hysteresis operators [2], [13], [14]. Thus, the mean-field model is one where two identical Preisach hysteresis operators are coupled to each other. The coupling constant $J$ corresponds to the strength of the competitive interaction. Local slope of hysteresis curves $d\hat{P}/dF$ produced by the decoupled Preisach operators is the probability density $\rho(F)$ of the coercive force distribution of the bi-stable switches. The maximum of the probability distribution $\rho_{max}$ is inversely proportional to its standard deviation and, therefore, the extent of the quenched-in disorder can be characterized by $\sigma = \frac{1}{\rho_{max}}$.

Using equations (2) it is easy to see that the stable states of the mean-field model can be alternatively described by the equations:

$$h = J\hat{P}[F_b] + F_a \quad (3a)$$

$$h = J\hat{P}[F_a] + F_b \quad (3b)$$

where $F_a = h - J\mu_a$ and $F_b = h - J\mu_b$ can be viewed as effective forces driving the evolution of each sub-group. Subtracting the above two equations from each other leads to the following equation:

$$(F_a - F_b) = (F_a - F_b) \cdot \frac{J}{2} \int_{-1}^{1} \rho \left[ \frac{(F_a + F_b)}{2} + \frac{\varepsilon}{2}(F_a - F_b) \right] d\varepsilon \qquad (4)$$

It is clear that when $J\rho_{max} = J/\sigma < 1$, the only solution of equation (4) corresponds to $F_a - F_b = 0$ and consequently $\mu_a - \mu_b = 0$. On the other hand, when $J\rho_{max} = J/\sigma > 1$, other solutions are possible where $F_a - F_b \neq 0$ and $\mu_a - \mu_b \neq 0$. Thus, by analogy with the classical mean-field model of equilibrium phase transition for antiferromagnets, quantity $\lambda = \mu_a - \mu_b$ can be viewed as the order parameter. Numerical implementation of the mean-field model described above gives the evolution of the order parameter with the external force along the increasing branch of the major hysteresis for different values of the tuning parameter $J/\sigma - 1 = r$ as shown in Figure 7.

The behavior of the order parameter illustrated in this figure above the critical value of the tuning parameter shows remarkable similarity to one that was obtained in simulations of the 2D RCIM (shown in Figure 6). This lends significant credibility to the results of our numerical simulations of the RCIM.

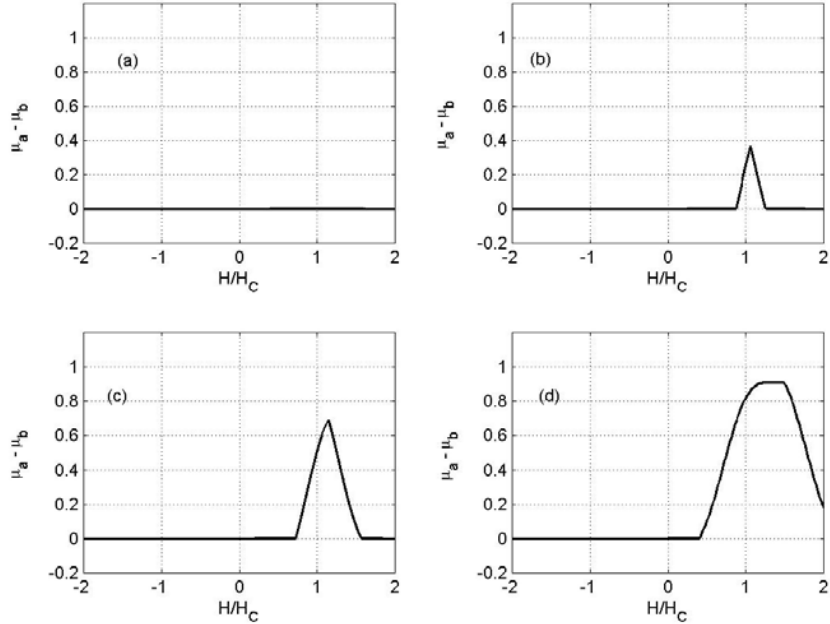

Figure 7 – Behavior of order parameter $\lambda = \mu_a - \mu_b$ at different tuning parameters a) $r \sim -0.5$, b) $r \sim 0.04$, c) $r \sim 0.2$ and d) $r \sim 1$

For the mean-field model, one can certainly study scaling of the order parameter with the tuning parameter $J/\sigma - 1 = r$ using equation:

$$1 = \frac{J}{2}\int_0^1 \rho\left[\frac{(F_a+F_b)}{2} + \lambda\frac{\varepsilon}{2}\right]d\varepsilon \qquad (5)$$

(that was obtained from equation (4)) and find that the classical mean-field scaling $\lambda \propto r^{0.5}$ holds near the coercive force where the maximum of the probability distribution occurs.

It can also be shown that RPM holds when $\lambda = 0$. Indeed, in this case $F_a = F_b = F$ making equation (2a) identical to (2b) and (3a) identical to (3b). Equations (3) and (2) respectively can now be written as:

$$h = J\hat{P}[F] + F \qquad (6a)$$

$$\mu = \mu_a = \mu_b = \hat{P}[F] \qquad (6b)$$

The first of the above relations implies that the external force is obtained from the internal force by a Preisach operator. The inverse of this operator exists because $\rho(F)$ is non-negative [14]. As a result, RPM is observed in this relation guaranteeing that closed cycles of $F$ produce closed cycles of $h$ and vice versa (closed cycles of $h$ produce closed cycles of $F$). The second relation similarly has RPM. As a result, the relation between $\mu$ and $h$ has to satisfy RPM. Any assumption to the contrary produces a contradiction. Thus, one can conclude that, for sufficiently low interaction strength or sufficiently high disorder, RPM will be observed.

Is the RPM preserved when the tuning parameter crosses the critical threshold? While we found no analytical way to prove that RPM is not preserved in this critical transition, we have performed numerical simulations of the mean-field model for different types of probability distribution functions $\rho(F)$. In all cases we observed that loss of RPM and onset of reptations is directly associated with this critical phenomenon. Figure 8 shows scaling of the range of reptations of the minor loops attached to the major hysteresis loop and Figure 9 shows scaling of minor loop opening at the reversal point integrated over the range of all possible minor loops. Both of these parameters show power law scaling with the tuning parameter. As expected, the exponents of these scaling relations are different then those obtained using numerical simulations of the 2D RCIM.

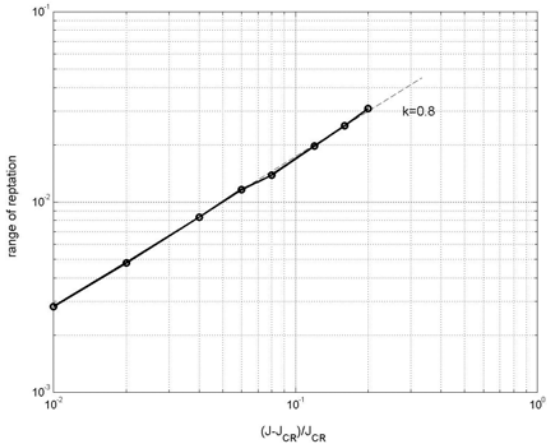
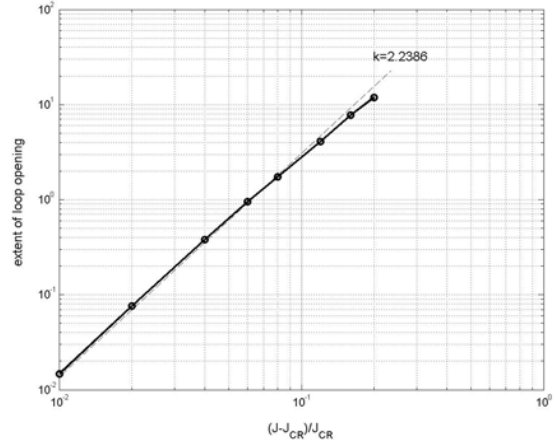

*Figure 8 – scaling of the range of reptations with tuning parameter for the mean-field model*

*Figure 9 – scaling of the extent of loop opening with the tuning parameter for the mean-field model.*

An analysis similar to the one carried out above for the RCIM can also be carried out for the RFIM with competitive interactions. In the mean-field approximation a similar type of critical phenomenon can be found. Interestingly, however, the mean-field model for RFIM consistently displayed RPM in numerical simulations both below and above the critical transition. This indicates that randomness in the coercive force is essential in describing an abrupt transition from RPM to reptations in hysteretic behavior.

CONCLUSION

Using the mean-field model we have found that a critical phenomenon occurs in systems with competitive interactions where local coercivity displays randomness. In such systems a state similar in structure to an antiferromagnetic ground state emerges near average coercivity when interactions are sufficiently large or the coercivity disorder is sufficiently small. Moreover, this critical phenomenon is associated with the loss of return point memory (RPM) and the onset of reptations in the minor hysteresis loops. Various measures describing the extent of reptations have been shown to scale with the tuning parameter according to power laws. While numerical experiments on systems with near-neighbor antiferromagnetic interactions indicated very similar behavior, confirmation of this type of critical phenomenon for these systems will require much larger system sizes and will be carried out in the near future.

REFERENCES

[1] G. Bertotti, *Hysteresis in Magnetism* (Academic Press, New York, 1998).
[2] I. D. Mayergoyz, Phys. Rev. Lett. **56**, 1518 (1986).
[3] E. Della Torre, F. Vajda, IEEE Trans Mag. **31**, 1775 (1995).
[4] J. S. Urbach, R. C. Madison, and J. T. Markert, Phys. Rev. Lett. **75**, 276, (1995)
[5] J. Guilmart, Z. Angew. Phys. **28**, 266 (1970)
[6] J.R. Petta, M.B. Weissman and G. Durin, Phys. Rev. **E56**, 2776 (1997);
    S. Zapperi and G. Durin, Comp. Mat. Sci. **20**, 436 (2001);
[7] L. Neel, Compt. rend. (Paris) **246**, 2313 (1958).
[8] L. Neel, J. Phys.radium **20**, 215 (1959).
[9] J. Sethna *et al*., Phys. Rev. Lett. **70**, 3347 (1993).
[10] T.G. Sorop, Phys. Rev. B **67**, 014402 (2003).
[11] K. Nielsch *et al*., Appl. Phys. Lett. **79**, 1360 (2001).
[12] A.H. Morrish, *The Physical Principles in Magnetism* (IEEE Press, New York, 2001).
[13] F. Preisach, Z. Phys. **94**, 277 (1935)
[14] M. Brokate, J. Sprekels, *Hysteresis and Phase Transitions* (Springer Verlag, New York, 1996).